\documentclass[twocolumn]{aastex62}

\received{xxxx, 2019}
\revised{xxxx, 2020}
\accepted{Octobar, 2020}

\shorttitle{Faint Quasars Live in the Same Number Density Environments as  Lyman Break Galaxies at $z\sim4$}
\shortauthors{Uchiyama et al.}

\begin{document} 
\title{Faint Quasars Live in the Same Number Density Environments as Lyman Break Galaxies at $z\sim4$} 
\correspondingauthor{Hisakazu Uchiyama}
\email{hisakazu.uchiyama@nao.ac.jp} 

\author{Hisakazu Uchiyama} 
\affiliation{National Astronomical Observatory of Japan, Mitaka, Tokyo 181-8588, Japan}  

\author{Masayuki Akiyama}  
\affiliation{Astronomical Institute, Tohoku University, Aramaki, Aoba, Sendai 980-8578, Japan}

\author{Jun Toshikawa} 
\affiliation{Institute for Cosmic Ray Research, The University of Tokyo, 5-1-5 Kashiwanoha, Kashiwa, Chiba 277-8582, Japan} 
\affiliation{Department of Physics, University of Bath, Claverton Down, Bath, BA2 7AY, UK}

\author{Nobunari Kashikawa}
\affiliation{Department of Astronomy, School of Science, The University of Tokyo, 7-3-1 Hongo, Bunkyo-ku, Tokyo, JAPAN, 113-0033}

\author{Roderik Overzier}
\affiliation{Observat\'orio Nacional, Rua Jos\'e Cristino, 77 CEP 20921-400, S\~ao Crist\'ov\~ao, Rio de Janeiro, Brazil}

\author{Tohru Nagao} 
\affiliation{Research Center for Space and Cosmic Evolution, Ehime University, Bunkyo-cho 2-5, Matsuyama 790-8577, Japan} 

\author{Kohei Ichikawa} 
\affiliation{Astronomical Institute, Tohoku University, Aramaki, Aoba, Sendai 980-8578, Japan} 

\author{Murilo Marinello} 
\affiliation{Laborat\'orio Nacional de Astrof\'isica, Rua dos Estados Unidos, N$^\circ$154, CEP37504-364 Itajub\'a, MG, Brazil} 

\author{Masatoshi Imanishi}
\affiliation{National Astronomical Observatory of Japan, Mitaka, Tokyo 181-8588, Japan } 
\affiliation{Department of Astronomical Science, The Graduate University for Advanced Studies, SOKENDAI, Mitaka, Tokyo 181-8588, Japan} 

\author{Masayuki Tanaka}
\affiliation{National Astronomical Observatory of Japan, Mitaka, Tokyo 181-8588, Japan } 
\affiliation{Department of Astronomical Science, The Graduate University for Advanced Studies, SOKENDAI, Mitaka, Tokyo 181-8588, Japan} 

\author{Yoshiki Matsuoka}
\affiliation{Research Center for Space and Cosmic Evolution, Ehime University, Bunkyo-cho 2-5, Matsuyama 790-8577, Japan }  

\author{Yutaka Komiyama} 
\affiliation{National Astronomical Observatory of Japan, Mitaka, Tokyo 181-8588, Japan} 

\author{Shogo Ishikawa}
\affiliation{Center for Computational Astrophysics, National Astronomical Observatory of Japan, Mitaka, Tokyo 181-8588, Japan}
\affiliation{National Astronomical Observatory of Japan, Mitaka, Tokyo 181-8588, Japan}

\author{Masafusa Onoue}
\affiliation{Max-Planck-Institut f\"{u}r Astronomie, K\"{o}nigstuhl 17, D-69117 Heidelberg, Germany}

\author{Mariko Kubo}
\affiliation{National Astronomical Observatory of Japan, Mitaka, Tokyo 181-8588, Japan} 

\author{Yuichi Harikane}
\affiliation{National Astronomical Observatory of Japan, Mitaka, Tokyo 181-8588, Japan} 
\affiliation{Department of Physics and Astronomy, University College London, London, WC1E 6BT, UK}

\author{Kei Ito}
\affiliation{Department of Astronomical Science, The Graduate University for Advanced Studies, SOKENDAI, Mitaka, Tokyo 181-8588, Japan}

\author{Shigeru Namiki}
\affiliation{Department of Astronomical Science, The Graduate University for Advanced Studies, SOKENDAI, Mitaka, Tokyo 181-8588, Japan}

\author{Yongming Liang}
\affiliation{Department of Astronomical Science, The Graduate University for Advanced Studies, SOKENDAI, Mitaka, Tokyo 181-8588, Japan}

\begin{abstract} 
Characterizing high-$z$ quasar environments is key to understanding the co-evolution of quasars and the surrounding galaxies. 
To restrict their global picture, 
we statistically examine the $g$-dropout galaxy overdensity distribution around 570 faint quasar candidates at $z\sim4$,  based on the Hyper Suprime-Cam Subaru Strategic Program survey. 
We compare the overdensity significances of $g$-dropout galaxies around the quasars with those around $g$-dropout galaxies, and find no significant difference between their distributions. 
A total of 4 (22) out of the 570 faint quasars, $0.7_{-0.4}^{+0.4}$ $ (3.9_{-0.8}^{+0.8})$\%, are found to be associated with the $>4\sigma$ overdense regions within an angular separation of $1.8$ (3.0) arcmin, which is the typical size of protoclusters at this epoch.  
This is similar to the fraction of $g$-dropout galaxies associated with the $>4\sigma$ overdense regions. 
This result is consistent with our previous work that $1.3_{-0.9}^{+0.9}$\% and  $2.0_{-1.1}^{+1.1}$\% of luminous quasars detected in the Sloan Digital Sky Survey exist in the $>4\sigma$ overdense regions within 1.8 and 3.0 arcmin separations, respectively. 
Therefore, we suggest that the galaxy number densities around quasars are independent of their luminosity, 
and most quasars do not preferentially appear in the richest protocluster regions at $z\sim4$. 
The lack of an apparent positive correlation between the quasars and the protoclusters  implies that: 
i) the gas-rich major merger rate is relatively low in the protocluster regions, ii) most high-$z$ quasars may appear through secular processes, or iii) some dust-obscured quasars exist in the protocluster regions.

\end{abstract}


\section{INTRODUCTION} 
High-$z$ quasars are powered by mass accretion onto supermassive black holes (SMBHs).  
According to the $M$-$\sigma$ relation \citep[][]{Magorrian98, Kormendy13}, 
the black hole mass is strongly correlated with the bulge mass, and thus, the halo mass of the host galaxy. 
If the local deterministic relation is valid even at a high redshift,  high-$z$ quasars are expected to reside in massive halos.  
In fact, at $z>3$, the quasar halo mass is massive: $M_{\text{h}} \!\sim\! (4 - 6) \times10^{12}~ h^{-1} M_{\odot}$ \citep[][]{Shen07}. 
In terms of the hierarchical structure formation, galaxies may frequently experience gas-rich mergers in galaxy overdense regions \citep[e.g.,][]{Hine16}. 
The central SMBH can be fed by sufficient gas to evolve into a quasar \citep[``merger scenario"; e.g.,][]{Hopkins08}.  
Thus, high-$z$ quasars are speculated to reside in galaxy overdense environments such as protoclusters. 

However, a consistent picture regarding the environments of $z>3$ quasars has not yet been achieved, despite numerous intensive studies \citep[For details, see ][]{Overzier16}. 
For instance, there are supportive evidences that $z>3$ quasars exist in galaxy overdense regions \citep[e.g.,][]{Zheng06, Kashikawa07, Utsumi10, Husband13, Morselli14, Balmaverde17, Decarli17, Garcia19, Neeleman19, Mignoli20}. 
On the other hand, several previous studies reported that quasars do not live in galaxy overdense regions at $z>3$ \citep[e.g.,][]{Toshikawa16, Mazzucchelli17, Kikuta17, Marino18, Ota18, Yoon19, Lupi19, Marinello19, Uchiyama19}.  
The characterization of high-$z$ quasar environments is challenging owing to the rarity of high-$z$ quasars and protoclusters. 
The number densities of protoclusters and quasars are expected to be $\sim1$ deg$^{-2}$ at $z\sim4$ \citep[][]{Toshikawa18, Uchiyama18}.  
Cosmic variance and the different physical-scale density measurement as well as the low statistics have prevented us from obtaining a global picture of the high-$z$ quasar environment. 

In recent years, \citet[][hereafter U18]{Uchiyama18} statistically examined whether 151 Sloan Digital Sky Survey (SDSS) quasars were correlated with 179 Hyper Suprime-Cam (HSC)-detected protoclusters at $z\sim4$ \citep[][]{Toshikawa18}, the number of which is approximately 10 times higher than previously known protoclusters. 
This more systematic study led to high statistics and low cosmic variance effects. 
As a result, the majority of the quasars at $z\sim4$ are likely not associated with protoclusters. 
Only $\sim1$\% of the SDSS luminous quasars were found to be associated with protoclusters.  
The SDSS quasar halos at $z\sim4$ were likely to evolve into group-like halos. 

The quasars used in U18, however, were limited to luminous ones with UV absolute magnitudes of $M_{\text{UV}} \lesssim -26$, corresponding to $\sim M_{\text{UV}}^{*}-0.5$ \citep[][]{Akiyama18} at $z\sim4$. 
The galaxy density around the less luminous quasars is likely to be higher than that around the luminous ones owing to  photoevaporation effects, which could make the surrounding galaxy densities lower \citep[e.g.][]{Kashikawa07, Uchiyama19, Dashyan19, Johnson19}. 
On the other hand, it is known that the clustering strength of quasars is almost independent of their luminosity \citep[e.g.,][]{Eftek15}, which means that the number density around less luminous quasars should be almost consistent with that around luminous quasars. 
To clarify whether less luminous quasars are located in protoclusters at $z\sim4$, we need to characterize the galaxy number density around each quasar statistically.    

In this study, we investigated the association between the galaxy overdensities (especially protoclusters) and 570 HSC-detected faint quasar candidates 
with a UV absolute magnitude of $M_{\text{UV}}=-26 \sim -22$ over a scale of $<3$ arcmin, which corresponds to the typical scale of protoclusters at $z\sim4$ \citep[][]{Chiang13}. 
By comparing the results of this study with those of U18,  
we could clarify the possible dependency on the quasar luminosity for their surrounding environments, and we could provide
more statistically robust results as our quasar sample size was approximately four times larger than that in the previous study of U18.  

The paper is organized as follows. 
In Section 2, we describe the HSC Subaru Strategic Program (HSC-SSP) survey, and the construction of the HSC-detected protoclusters and quasars. 
In Section 3, we investigate the environments of the HSC quasars. 
The implications of our results are discussed in Section 4. Finally, in Section 5, conclusions and a summary of our findings are provided. 
We assume the following cosmological parameters: 
$\Omega_{M} = 0.3 $, $\Omega_{\Lambda} = 0.7$, $H_{0} = 70~ $km s$^{-1}$ Mpc$^{-1}$, and magnitudes are given in the AB system.

\section{HSC-SSP SURVEY AND SAMPLE SELECTION}
\subsection{Subaru HSC-SSP survey}
The HSC-SSP survey started in early 2014 and will include 300 nights until its completion by 2020. 
The HSC is equipped with 116 2K $\times$ 4K Hamamatsu fully depleted CCDs, among which 104 CCDs are used to obtain scientific data over a field-of-view of 1.$^\circ$5 in diameter. The detailed system design is summarized by \citet{Miyazaki18}. 
For the CCD dewar and camera system design, one should refer to \citet{Komiyama18}.  
This study is based on the Wide layer of the HSC-SSP, which has wide-area coverage and high sensitivity through five optical bands ($g$, $r$, $i$, $z$, and $y$) with transmission functions \citep[][]{Kawanomoto18} that are roughly similar to the SDSS filter curves. 
The total exposure times are $10$ min in the $g$ and $r$ bands and $20$ min in the $i$, $z$, and $y$ bands. 
The on-site quality assurance system for the HSC \citep[OSQAH][]{Furusawa18} is used in the HSC-SSP observations. 
The HSC data \citep[DR S16A][]{Aihara18a} in the Wide layer have already produced an extremely wide field image of $>200$ deg$^2$ with a median seeing value of $0.\arcsec6 - 0.\arcsec8$. 
The survey design is presented in \citet{Aihara18b}.  
Data reduction is performed with the dedicated pipeline hscPipe \citep[version 4.0.2, ][]{Bosch18}, which is a modified version of the Large Synoptic Survey Telescope software stack \citep[][]{Ivezic08, Axelrod10, Juric15}. 
The astrometric and photometric calibrations are associated with the Pan-STARRS1 system \citep[][]{Schlafly12, Tonry12, Magnier13}. 
The fluxes and colors of the sources are measured with cModel, which is achieved by fitting two-component, point spread function (PSF)-convolved galaxy models to the source profile \citep[]{Abazajian04}.

\subsection{HSC protocluster sample}
We use the $z\!\sim\!4$ protocluster candidate catalog constructed from DR S16A \citep[][]{Toshikawa18}. 
We briefly summarize the key steps of the catalog construction here. 

\begin{figure}
\begin{flushleft}
 \includegraphics[width=1.0\linewidth]{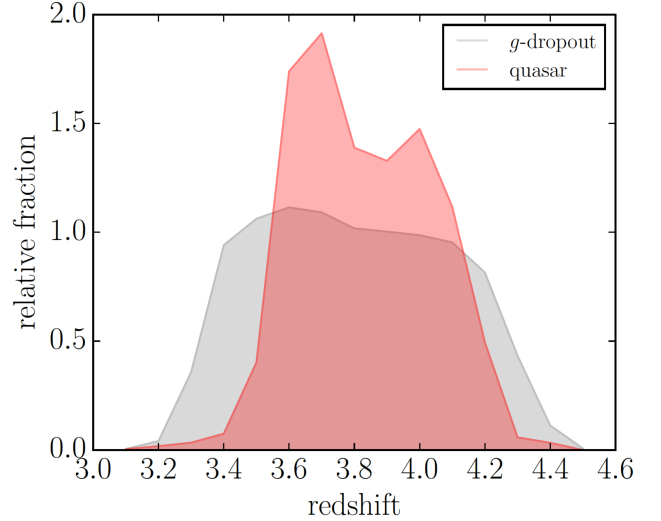}
\end{flushleft}
\caption{Redshift distributions of $g$-dropout galaxy \citep[gray shade; ][]{Ono18} and HSC quasar \citep[red shade; ][]{Akiyama18}. The distributions are normalized by $\int dz f(z)$, where $f(z)$ is each redshift distribution function. 
}\label{sf}
\end{figure}

Firstly, the $g$-dropout galaxies were selected from five independent fields: W-XMMLSS (R.A.$=1^{h}36^{m}00^{s}$ $\sim$ $3^{h}00^{m}00^{s}$, Decl. =  $-6^{\circ}00'00''$ $\sim$ $-2^{\circ}00'00''$), W-Wide12H (R.A. $=11^{h}40^{m}00^{s}$ $\sim$ $12^{h}20^{m}00^{s} $, Decl. $=-2^{\circ}00'00'' $  $\sim$  $2^{\circ}00'00''$), W-GAMA15H (R.A. $=14^{h}00^{m}00^{s}$ $\sim$ $15^{h}00^{m}00^{s} $, Decl. $=  -2^{\circ}00'00'' $ $\sim$ $ 2^{\circ}00'00'' $), W-HECTOMAP (R.A. $=15^{h}00^{m}00^{s}$ $\sim$ $17^{h}00^{m}00^{s} $, Decl. $= 42^{\circ}00'00'' $ $\sim$  $45^{\circ}00'00''$), and W-VVDS (R.A. $=22^{h}00^{m}00^{s}$ $\sim$ $23^{h}20^{m}00^{s} $, Decl. $=-2^{\circ}00'00'' $ $\sim$ $ 3^{\circ}00'00'' $).  
Here, the following criteria were used, as provided in \citet{vanderBurg10} : 
\begin{eqnarray}
  1.0  & < & g-r,  \label{eq1}\\ 
          r-i &  < &1.0, \\
1.5 (r-i) & < &(g-r) -0.80, \\
 r  & \le & m_{\mathrm{lim}, 3\sigma}, \\ 
  i  & \le & m_{\mathrm{lim}, 5\sigma},  \label{eq5}
\end{eqnarray}
where $m_{\mathrm{lim}, 3\sigma}$ and $m_{\mathrm{lim}, 5\sigma}$ are $3\sigma$ and $5\sigma$ limiting magnitudes, respectively. 
If the objects were not detected in the $g$-band filter at 3$\sigma$, their $g$-band magnitudes were replaced with the corresponding 3$\sigma$ limiting magnitudes. 
The red galaxies at intermediate redshifts and dwarf stars could satisfy the color selection criteria. 
The contamination rate from these objects was evaluated as a maximum of 25\% at $i<25.0$ \citep[]{Ono18}.  
The $g$-dropout selection completeness window was estimated to be $z=3.8_{-0.5}^{+0.5}$ by the Monte Carlo method \citep[]{Ono18}, as illustrated in Figure \ref{sf}.  

Owing to the sky conditions in the observation, the depth over the entire regions is inhomogeneous, 
which causes large scatter in the number of detected objects in the different fields, leading to difficulties in comparing the number densities among their fields. 
Thus, only homogeneous depth regions were used, where the $g$-, $r$-, and $i$-band $5\sigma$ limiting magnitudes were fainter than 26.0, 25.5, and 25.5 mag, respectively. 
The resultant effective area after removing the masked region around the bright objects was $121$ deg$^2$. 
In this area, 259,755 $g$-dropouts down to $i<25.0$ were detected.

Subsequently, the fixed aperture method was applied to determine the surface density contour maps of the $g$-dropout galaxies. 
Apertures with a radius of $1.8$ arcmin, corresponding to $0.75$ physical Mpc (pMpc) at $ z \!\sim\! 4$, were distributed in the sky of the HSC Wide layer. 
This aperture size is comparable with the typical protocluster size at this epoch with a descendant halo mass of $\gtrsim 10^{14}~ M_{\odot}$ at $z=0$ \citep[]{Chiang13}. 
The resolution of the density map was $1\times1$ arcmin$^2$. 
The overdensity significance was defined by ($N$-$\bar{N}$)/$\sigma$, where $N$ is the number of $g$-dropout galaxies in an aperture, 
and $\bar{N}$ and $\sigma$ are the average and standard deviation of $N$ over the entire fields, respectively. 
($\bar{N}$, $\sigma$) was estimated to be ($6.4$, $3.2$) \citep[][]{Toshikawa18}. 
It was assumed that the number density of the $g$-dropout galaxies in the masked regions was the same as the average.  
However, the apertures in which $>50$\% of the area was masked were excluded. 
As the contaminants of foreground objects distribute randomly over the sky, the overdensity significance should not be greatly affected by the contamination \citep[][]{Toshikawa12}.

The protocluster candidates were defined as overdense regions with $>4\sigma$ overdensity significance. 
Although large scatter exists owing to projection effects, the surface overdensity significance is closely correlated with the descendant halo mass at $z = 0$. 
\citet{Toshikawa16} demonstrated that $ > 76 \%$ of $> 4\sigma$ overdense regions of $g$-dropouts were expected to evolve into dark matter halos with masses of $> 10^{14}~ M_{\odot} $ at $z=0$, 
whereas overdense regions with less than $4\sigma$ significance could not be distinguished from fields owing to the projection effect.  
According to \citet{Toshikawa18}, as a result of the projection effect, the protocluster sample is biased to the richest structure, the average descendant halo mass of which is expected to be $\sim5\times10^{14} M_{\odot}$, with a high purity ($>76\%$) but low completeness ($\sim6$\%). 
By clustering analysis, their mean halo mass at $z\sim4$ was estimated to be $2.3_{-0.5}^{+0.5}\times10^{13} h^{-1} M_\odot$, which is comparable to galaxy group halo mass \citep[][]{Toshikawa18}. 
It should be noted that the success rate of this technique has already been established by a previous study on the $\!\sim\! 4$  deg$^2$ of the CFHTLS Deep Fields, as the precursor of this HSC protocluster research, followed by Keck/DEIMOS and Subaru/FOCAS spectroscopy \citep[]{Toshikawa16}. 
The authors carefully checked each $>4\sigma$ overdense region, and removed $22$ fake detections of mainly the spiral arms of local galaxies \citep[][]{Aihara18a}. 
As a result, $179$ protocluster candidates at $z \!\sim\! 4$ in the HSC Wide layer with overdensity significance ranging from $4$ to $10\sigma$ were obtained.

\begin{figure*}
\begin{flushleft}
 \includegraphics[width=1.0\linewidth]{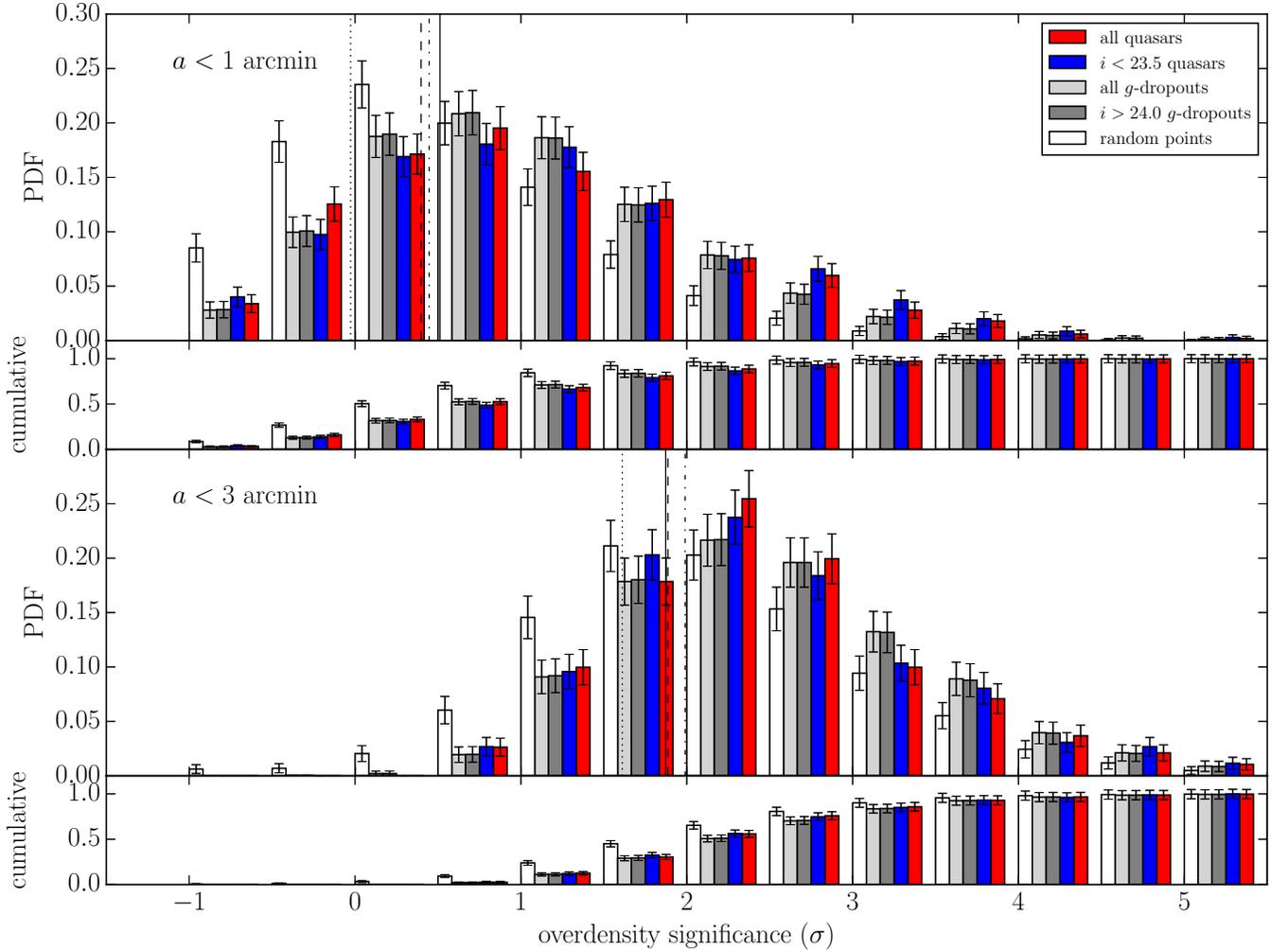}
\end{flushleft}
\caption{
The distributions of the local maxima of the overdensity significances within radii of 1 and 3 arcmin centered on HSC quasars (red bars), bright HSC quasars with $i<23.5$ (blue bars), $g$-dropout galaxies (light gray bars), $g$-dropout galaxies  with $i>24.0$ (gray bars), and random points (white bars) are illustrated in the upper and lower panels, respectively. 
The error bar assumes Poisson error. 
The vertical dashed, solid, dotted, and dashed-dotted lines represent the median values of the overdensity significance distribution of the HSC quasars, bright HSC quasars, random points, and $g$-dropout galaxies, respectively. 
}\label{overdensity} 
\end{figure*}

\begin{figure*}
\begin{center}
  \includegraphics[width=1.0\linewidth]{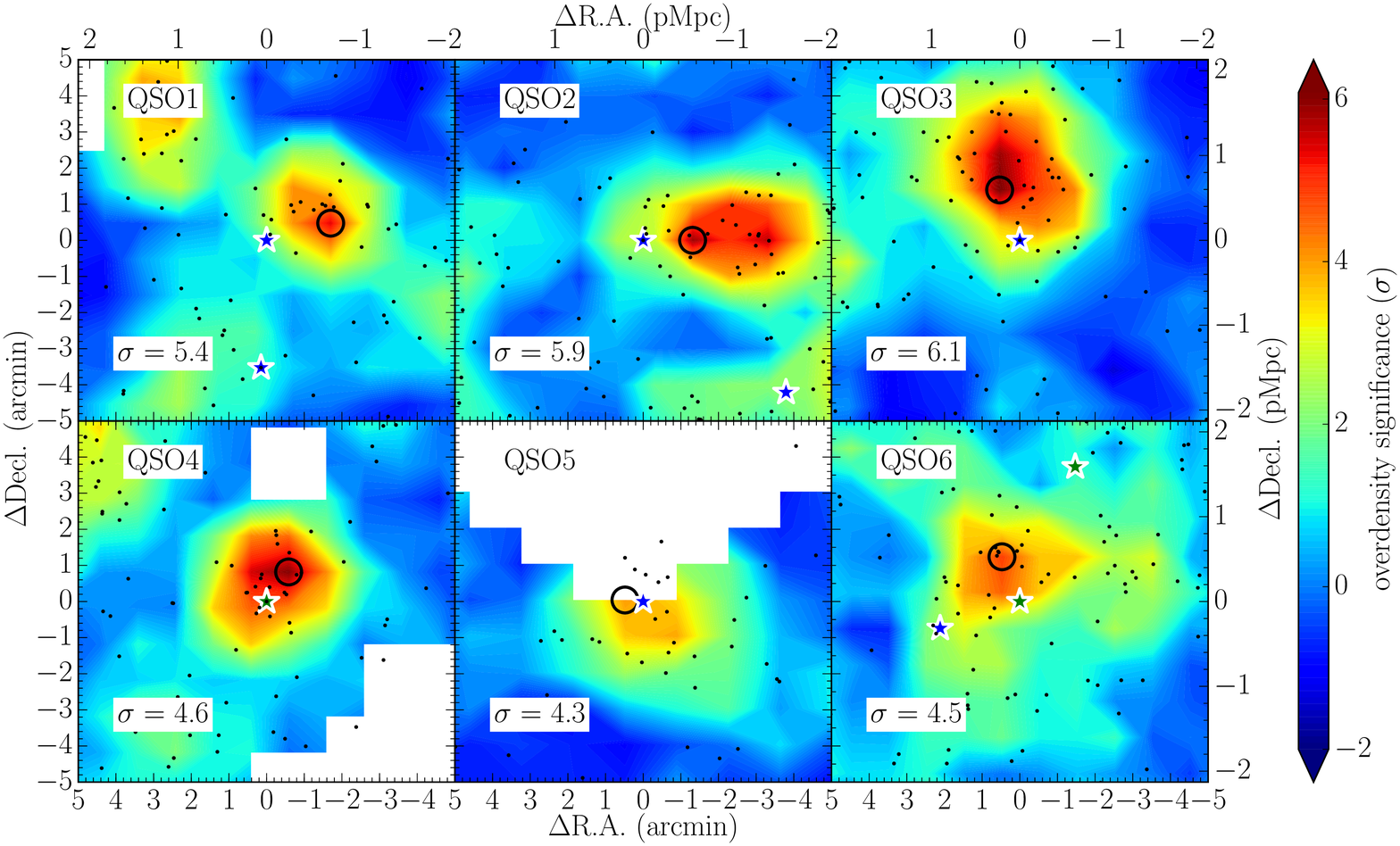}
\end{center}
\caption{Overdensity maps around $z\sim4$ quasars that could associate with protocluster candidates. 
The quasars with ID $=$ QSO1$-$6 are centered at each panel. 
The blue and green star symbols indicate the HSC and SDSS quasars, respectively, that reside in the regions within 1.8 arcmin ($0.75$ pMpc) of the protocluster candidates. 
The peaks of the overdensity significances are denoted by the black circles. 
These peak values are indicated in the lower left corner of each panel. 
The overdensity significance is represented by the color contours. 
The black dots indicate the $g$-dropout galaxies and the regions that are not used in the protocluster search owing to bright objects (section 2.2) are indicated by the white regions. 
The size of each panel is $10$ arcmin $\times$ $10$ arcmin. 
}\label{map}
\end{figure*}

\subsection{HSC quasar sample} 
The $z\sim4$ quasar sample was constructed from the same DR S16A as used in \citet{Akiyama18}.  
We present the essence of the selection below. 

The authors selected the quasar sample conservatively in the effective survey area of 172.0 deg$^2$, which is covered by $grizy$-bands and masked around bright objects ($i<22$) 
to avoid possible failure of the HSC pipeline de-blending process for faint objects in the outskirts of the bright objects \citep[][]{Akiyama18}. 
Stellar objects were extracted when their second-order adaptive moments \citep[][]{Hirata03} were comparable to the PSF size. 

The $z\sim4$ quasar candidates were selected by using the $grz$-band color--color selection, 
\begin{eqnarray}
  r-z  ~& < &~ 0.65 ~(g-r) -0.30,  \label{eq1}\\ 
          r-z ~ &  < & ~3.50 ~(g-r) - 2.90 , \\
g-r ~& < &~1.50, \label{eq10}
\end{eqnarray}
and  $izy$-band color--color selection, 
\begin{eqnarray}
z-y  ~& < &~ -2.25 ~(i-z)+0.400,  \label{eq1}\\  
          -0.3 ~&  < & ~ i-z. \label{eq15}
\end{eqnarray}
The latter $izy$-band criterion was applied to remove contaminations such as red galactic stars and several outliers. 
\citet{Akiyama18} limited the sample to $i> 20$ because bright objects ($i<20$) are affected by saturation or non-linearity effects.  
They carefully removed objects around bright stars and galaxies, as well as those close to faint halos and satellite tracks around bright stars to prevent unreliable photometry.

A total of 1,668 quasar candidates, with photometric redshifts of $z=3.9_{-0.3}^{+0.4}$ (Figure \ref{sf}), were detected. 
The UV absolute magnitudes of the HSC quasars lie in the range of $-26$ to $-22$. 
The faintest UV magnitude is $\sim4$ mag fainter than that of the SDSS quasars selected at $z\sim4$ by \citet{Akiyama18}. 
The contamination rate of the HSC quasars is $<40$\% at $i<23.5$, and $>50$\% at $i>23.5$. 
We confirm that our results do not change, even if the sample is limited to bright quasars with $i<23.5$ (Section 3). 
The completeness is higher than 80\% at $i<23.0$ and decreases to $\sim40$\% at $i=24$. 
Note that the survey area is wider than that of our $g$-dropout galaxies. 

\subsection{Cross-matched effective area}
We cross-match the effective areas used in the protocluster and quasar searches to examine their possible correlation fairly. 
As a result, a cross-matched effective area of $\sim98$ deg$^2$ is established. 
In total, $146$ protoclusters ($210,627$ $g$-dropout galaxies with $i<25.0$) and $570$ HSC quasars in the effective area are obtained. 
As the redshift distribution of the HSC quasars is slightly narrower than that of the $g$-dropout galaxies, as illustrated in Figure \ref{sf}, we aim to determine not whether the $g$-dropout overdense regions host the HSC quasars, but whether the number density environments around the HSC quasars are overdense regions.   

\section{RESULTS}

\begin{deluxetable}{lllll}[t!]
\tablecaption{List of protoclusters associated with quasars \label{t1}}
\tablecolumns{7}
\tablenum{1}
\tablewidth{0pt}
\tablehead{
 \colhead{ Name \tablenotemark{a}} & \colhead{R.A. (J2000)} & \colhead{Decl. (J2000)} & \colhead{$\Sigma$ \tablenotemark{b}} & \colhead{Separation \tablenotemark{c}} \\
 \colhead{}           & \colhead{}             & \colhead{}              & \colhead{($\sigma$)}                 & \colhead{(pMpc)} 
}
\startdata
 QSO1          &  $02^{h}15^{m}23.13^{s}$ & $-02^{\circ}40'36.1'' $ &  5.4 &  0.74 \\ 
 QSO2          &  $02^{h}18^{m}53.93^{s}$ & $-05^{\circ}33'07.6'' $ &  5.9 &  0.54  \\ 
 QSO3          &  $02^{h}25^{m}27.23^{s}$ & $-04^{\circ}26'31.2'' $ &  6.1 &  0.62 \\ 
 QSO4          &  $14^{h}47^{m}13.04^{s}$ & $-01^{\circ}21'58.6'' $ &  4.6 &  0.42 \\ 
 QSO5          &  $15^{h}56^{m}22.98^{s}$ & $+44^{\circ}16'57.5'' $ &  4.3 &  0.15 \\ 
 QSO6          &  $22^{h}14^{m}58.38^{s}$ & $+01^{\circ}07'36.1'' $ &  4.5 &  0.55 
\enddata
\tablenotetext{a}{QSO4 and QSO6 are SDSS quasars, and the others are HSC quasars.  
}
\tablenotetext{b}{
Peak significance of protocluster.
}
\tablenotetext{c}{
Separation from quasar to nearest protoclusters. 
}
\end{deluxetable}

\subsection{Overdensity distribution around quasars}

We derive the distribution of the local maxima of the overdensity significances within a typical protocluster radius centered on the HSC quasars. 
This simple approach is sufficient to understand whether or not the quasars reside in the protoclusters. 
Note that in this approach, we cannot assess the possible local clustering signal within $<1.8$ arcmin radius from the quasars, as the overdensity significances are  calculated by using apertures with a radius of $1.8$ arcmin. 

The distributions of the maximum overdensity significances within circles with radii of $a = 1$ arcmin centered on the quasars and random points are illustrated in the upper panel of Figure \ref{overdensity}. 
Here, the random points are 570 points distributed uniformly on the cross-matched effective area, and 
their overdensity significance distribution is estimated by taking the average of 10,000 realizations. 
We find that the HSC quasars tend to reside in denser regions than the random points. 
In fact, the $P$-value in the Kolmogorov--Smirnov (KS) test for their distributions is estimated to be $\sim0$. 

To obtain a quantitative understanding of the number density-based environment of the HSC quasars, 
we also compare the overdensity distributions around the HSC quasars with those around the $g$-dropout galaxies. 
The probability distribution of the maximum overdensity significances within the $a = 1$ arcmin radius centered on the $g$-dropout galaxies is also depicted in the upper panel of Figure \ref{overdensity}. 
The overdensity significance distribution of the quasars is found to be significantly the same as that of the $g$-dropout galaxies.  
The KS test $P$-value for them is $0.09$. 
This result implies that the clustering strength of quasars and $g$-dropout galaxies is similar at least at the scales probed herein (i.e., 1.8 and 3 arcmins).

As shown in Figure 3, we also find that only four out of 570 ($=0.7_{-0.4}^{+0.4}$\%, where the error is assumed as Poisson error) HSC quasars are associated with the $>4\sigma$ overdense regions within the 1.8 arcmin separation \footnote{ 
Additionally, $2.3_{-0.6}^{+0.6} (10_{-1.3}^{+1.3})$\% of the HSC quasars, are associated with the overdense regions with $>3\sigma$ ($>2\sigma$) overdensity significances within 1.8 arcmin separation.}. 
The positions of the protoclusters associated with the quasars are summarized in Table \ref{t1}. 
Among the 570 randomly selected $g$-dropout galaxies, the fraction of those associated with the $>4\sigma$ overdense regions within the separation radius of $1.8$ arcmin is $0.5_{-0.3}^{+0.3}$\%. 
Here, the error is estimated by random sampling with replacement of the 570 $g$-dropout galaxies 10,000 times (bootstrap method). 
As a result, this ratio is the same as that of the HSC quasars within $1\sigma$ error. 
This suggests that the HSC quasars do not preferentially appear in the richest protoclusters. 

Note that the radial size of the protocluster at $z\sim4$ could extend to a maximum of $\sim3.0$ arcmin \citep[][]{Chiang13}. 
Even if we use a radius of $a = 3$ arcmin ($= 1.2$ pMpc) to evaluate the local maximum distribution of the overdensity significances, no significant difference of the overdensity significance distributions between the quasars and the $g$-dropout galaxies is found, as indicated in the lower panel of Figure \ref{overdensity}. 
The KS test $P$-value for their distributions is 0.15. 
Moreover, only 22 out of 570 ($=3.9_{-0.8}^{+0.8}$\%) quasars reside in the $>4\sigma$ overdense regions\footnote{  
$10_{-1.3}^{+1.3} (30_{-2.3}^{+2.3})$\% of the HSC quasars are associated with the $>3\sigma$ ($>2\sigma$) overdense regions within 3.0 arcmin separation. 
}. 
Again, the ratio is consistent with that of the $g$-dropout galaxies within $1\sigma$ error; Among the 570 randomly selected $g$-dropout galaxies, the fraction of those associated with the protoclusters within the $3.0$ arcmin separation is $4.4_{-0.9}^{+0.9}$\%. 
The results do not change even if the HSC quasar sample is limited to $i<23.5$ bright ones as shown in Figure \ref{overdensity}. 

In our analysis, we can ignore quasar contaminants in the $g$-dropout galaxy sample as the quasar fraction of the $g$-dropout galaxies is only $\sim0.01$. 
Here, this value is estimated by using the $g$-dropout luminosity function, $f$, and the galaxy fraction, $w$, given by \citep[][]{Ono18}, and taking a weighted-average of ($1-f$) with weight $w$ on a magnitude range of $i<25$.  
In fact, even if we use a subsample of the faintest $g$-dropout galaxies with $i>24$, where the quasar  contamination is much lower \citep[][]{Ono18}, to measure the overdensity distribution, our results do not change (Figure \ref{overdensity}).

\subsection{Stacking overdensity maps around quasars}  
\begin{figure}
\begin{center}
 \includegraphics[width=1.0\linewidth]{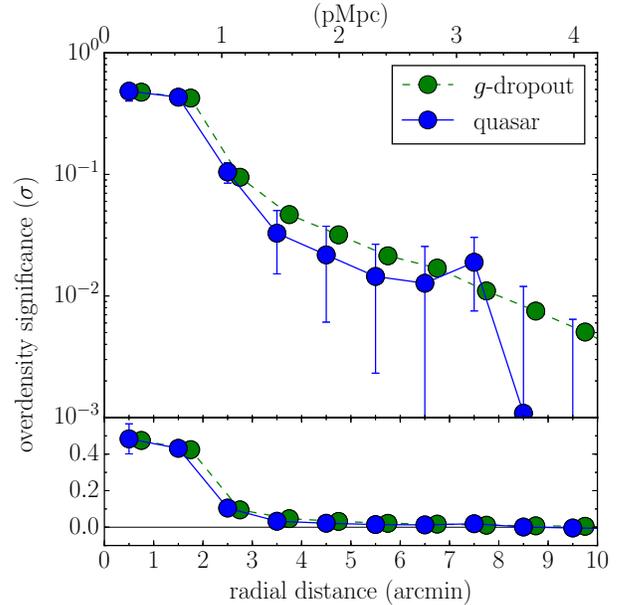} 
\end{center}
\caption{Stacked radial profile of overdensity significance distribution centered on HSC quasars/$g$-dropouts. 
The blue points indicate the stacked overdensity profile of the HSC quasars. 
The green points, which are offset by $+0.25$ arcmin over the horizontal axis for improved visualization, indicate those of all $g$-dropout galaxies. 
The error bars represent the $95$\% confidence interval based on the standard error of the mean of the overdensity significances in each bin. 
The logarithmic and the linear scale of the vertical axis are shown in top and bottom panels respectively. 
}\label{stack}
\end{figure}

We stack overdensity significance maps around the HSC quasars and the $g$-dropout galaxies with a radial scale of $10$ arcmin ($\sim4$ pMpc) to examine the possible difference in the galaxy density environments between the HSC quasars and $g$-dropout galaxies at any scale. 
Figure \ref{stack} presents the radial profiles of the stacked overdensity significance maps centered on the HSC quasars and the $g$-dropout galaxies. 
For a fair comparison of their density environments, we exclude the 570 HSC quasars from the $g$-dropout galaxy sample. 
We find that the galaxy density profile around the HSC quasars is statistically identical to that of the $g$-dropout galaxies at all scales within a 95\% confidence interval.  

\section{DISCUSSION} 
\subsection{Environment and halo mass of quasar at $z\sim4$}
We found that 4 (22) out of the 570 HSC quasars; that is, $0.7_{-0.4}^{+0.4}$ $(3.9_{-0.8}^{+0.8})$\% of the HSC quasars, are associated with the overdense regions with $>4\sigma$ overdensity significances within 1.8 (3.0) arcmin separation at $z\sim4$ (but see the caveat explained in Section 4.2). 
Now, we need to confirm that this ratio is consistent with the perspective of the quasar halo mass. 

We use the extended Press--Schechter (EPS) model \citep[][]{Press74, Bower91, Bond91, Lacey93} to examine whether or not the quasar halos at $z\sim4$ evolve into the cluster halos at $z\sim0$, or, in other words, their quasars can belong to protoclusters at $z\sim4$. 
According to the EPS model, the conditional probability $P_2(M_{t2}, z_2 | M_{t1}, z_1)$ that a halo with a mass of $M_{t1}$ at a redshift of $z_1$ evolves into a halo with a mass of $M_{t2}$ at a later redshift of $z_2$ is estimated as follows \citep[][]{Hamana06}: 
\begin{eqnarray} 
&& P_2(M_{t2}, z_2 | M_{t1}, z_1) dM_{t2}  \nonumber \\
&=& \frac{1}{\sqrt{2\pi}} \frac{\delta_{c2} (\delta_{c1} - \delta_{c2})}{\delta_{c1}} \left[ \frac{\sigma_1^2}{\sigma_2^2 (\sigma_1^2 - \sigma_2^2)} \right]^{\frac{3}{2}} \nonumber \\
&\times& \text{exp}\left[ -\frac{(\sigma_2^2 \delta_{c1} - \sigma_1^2 \delta_{c2})^2}{2\sigma_1^2 \sigma_2^2 (\sigma_1^2 - \sigma_2^2)} \right] \left| \frac{d \sigma_2^2 }{d M_{t2}} \right| d M_{t2}, \label{eq11} 
\end{eqnarray}
where $\sigma_i$ and $\delta_{ci}$ are the root mean square density fluctuation smoothed over spheres of a mass of $M_{ti}$ at $z_i$ and the critical overdensity threshold to collapse at $z_i$, respectively.  
Subsequently, the conditional mass function $n_2(M_{t2}, z_2 | M_{t1}, z_1)$ is estimated from equation (\ref{eq11}) \citep[][]{Hamana06}: 
\begin{eqnarray}
&& n_2(M_{t2}, z_2 | M_{t1}, z_1) dM_{t2}  \nonumber \\
&\propto& \frac{1}{M_{t2}}   P_2(M_{t2}, z_2 | M_{t1}, z_1) dM_{t2}. \label{eq12} 
\end{eqnarray}
\citet{He18} estimated the average halo masses of $1.46_{-1.01}^{+1.71} \times10^{12} h^{-1} M_\odot$ for the HSC quasars at $z=3.9_{-0.3}^{+0.4}$ by means of clustering analysis. 
We set $M_{t1}=1.46_{-1.01}^{+1.71} \times10^{12} h^{-1} M_\odot$, $z_1 = 3.9_{-0.3}^{+0.4}$, and $z_2 = 0.0$. 
Thereafter, the halo mass distribution at $z=0$ can be calculated from equations (\ref{eq11}) and (\ref{eq12}). 
We find that $8.4_{-6.6}^{+14.6}$\% of the halos with a mass of $1.46_{-1.01}^{+1.71} \times10^{12} h^{-1} M_\odot$ at $z = 3.9$ are expected to evolve into halos with a mass of $>10^{14} M_\odot$ at $z=0$.  
The ratio is almost the same, with $1.2\sigma$ ($0.7\sigma$) significance, as that of the HSC quasars associated with the $>4\sigma$ overdense regions out of the entire sample within 1.8 (3.0) arcmin separation. 
This means that our findings are consistent with the expected halo mass growth. 

\subsection{Signal dilution over redshift space}
Our overdense region sample exhibits high incompleteness owing to the large redshift error of the $g$-dropout galaxies (Section 2.2). 
We quantitatively evaluate to what extent the possible clustering signals around the HSC quasars are diluted by this projection effect. 

A quasar is assumed to exist in a typical protocluster-scale cylinder, with a diameter of $1.5 (1+z)$ comoving Mpc (cMpc) along the celestial sphere direction and a height of $1.5 (1+z)$ cMpc along the line-of-sight (LoS) direction at $z=3.9$ which corresponds to the median redshift of the HSC quasars.  
The cylinder space is assumed to be an overdense region with $n$ times the average galaxy density (cMpc$^{-3}$) at the cylinder scale. 
By integrating the $g$-dropout selection function (Figure \ref{sf}), the redshift window width of the $g$-dropout galaxy is estimated to be $>10$ times the cylinder scale (protocluster scale) along the LoS direction. 
Thus, a cylinder with $n>20$ times the average density can be achieved for our $>4\sigma$ overdense regions, using the definition of the overdensity significance ($N$-$\bar{N}$)/$\sigma$ (Section 2.2). 
Conversely, if the vicinity of a quasar is an overdense region with $<20$ times the average density, the density region is not detected as our protocluster sample.  
This implies that few $>4\sigma$ overdense regions (the most prominent ones) are detected (the completeness is $\sim6$ \%; see Section 2.2), and therefore, a higher fraction of quasars may be associated with protoclusters. 

\subsection{Comparison with SDSS luminous quasars}  
In order to examine the possible dependency of the quasar environments on their luminosities, we compare our findings with the results of U18, in which the $g$-dropout galaxy number density environments of luminous quasars at $z\sim4$ were investigated by using HSC-detected protoclusters. 
The method for selecting the $g$-dropout galaxies and the protoclusters were the same as in the present study. 
The quasars were extracted from the SDSS quasar catalog based on SDSS DR12 \citep[][]{Paris17}.  
The SDSS DR12 quasar catalog is the final SDSS-III quasar catalog, which contains $297,301$ spectroscopically confirmed quasars over a wide redshift range of $0.041 < z < 6.440$ in an area covering approximately $10,200$ deg$^2$ of the sky. 
A total of $151$ SDSS quasars are found in the effective areas of the HSC protocluster search within the $g$-dropout redshift range of $z=3.3-4.2$.  
The UV absolute magnitudes of this sample are brighter than $\sim-26$. 
For details, please refer to U18.  

We find that our results in the present study support our previous study of U18; 
2 and 3 out of 151 SDSS quasars ($1.3_{-0.9}^{+0.9}$\% and $2.0_{-1.1}^{+1.1}$\%, where these errors are  assumed to be Poisson errors) are found to be spatially associated with the $>4 \sigma$ overdense regions within 1.8 and 3.0 arcmin separations, respectively.  
Their ratios are consistent, within a $1\sigma$ error, with the case of the HSC quasars. 
This fact suggests that the number densities of the $g$-dropouts around the quasars are expected to be  independent of their luminosity over a wide range of their UV absolute magnitudes, $M_{\text{UV}}=-28\sim-22$. 
This may arise from the fact that the halo mass is almost independent of the quasar luminosity \citep[e.g.,][]{Eftek15}. 
In fact, \citet{He18} found that the halo masses of the HSC quasars were consistent with those of the SDSS quasars\footnote{For example, using the maximum likelihood method fitting to a power-law model, the halo masses of the HSC quasars and the SDSS quasars were estimated to be log $M_h/h^{-1} M_\odot = 12.20_{-0.49}^{+0.33}$ and log $M_h/h^{-1} M_\odot = 11.43_{-3.00}^{+0.88}$, respectively. For details, please refer to \citet{He18}. }.

\subsection{Environment of quasar pairs at $z\sim4$}
As illustrated in Figure \ref{map}, two out of the four HSC quasars associated with the $>4\sigma$ overdense regions, namely QSO1 and QSO2, could constitute possible quasar pairs with their projected separations of 9.6 arcmin ($=4$ pMpc). 
In \citet{Onoue18a}, a quasar pair was defined as two quasars with their separation closer than the size of massive protoclusters that are expected to evolve into the descendant halos with halo masses of $10^{15} M_\odot$ at $z=0$. 
Concretely, they extracted the quasar pairs with their projected proper distance of $R_\perp<4$ pMpc and velocity offset of $\Delta V < 3000$ km s$^{-1}$. 
We define the two quasars with their projected proper distance $R_\perp<4$ pMpc as the ``possible quasar pairs'', as we have no information regarding the velocity offset in the HSC quasar sample.  
Our finding is likely to relate to the results found in the SDSS quasars by \citet{Onoue18a}, whereby the SDSS quasar pairs tended to reside in the galaxy overdense regions at $z\sim4$. 
Spectroscopy is required for our possible quasar pairs to confirm the reality of the quasars and to measure their redshift differences.

\subsection{Triggering mechanism of quasar}
Our findings show no preference of quasar activity in the $z\sim4$ richest protocluster environment, where gas-rich galaxy major mergers are expected to occur frequently \citep[e.g.,][]{Gottlober01, Stewart09, Jian12, Hine16}. 
This fact is likely to be against the simple expectation that almost all of gas-rich major mergers result in quasars. 
We show some implications for our results below. 

First, the major merger rates in our protocluster regions may be not so high that we cannot confirm the enhancement in our statistical accuracy. 
The projection effect discussed in Section 4.2 prevents the detection of the possible enhancement.  
The major merger rates can be diluted to be $<0.1$ times owing to the effect. 
Using Hubble Space Telescope imaging combined with spectroscopic observation for the quasar-hosted protoclusters is necessary to estimate the exact merger rates \citep[][]{Hine16}. 


Second, if the major mergers occur with non-negligible probability in the protocluster regions, they may not activate SMBHs effectively. 
Almost all of the high-$z$ quasars may appear through not the major mergers but secular processes \citep[e.g.,][]{Fanidakis12}. 
Several recent observations have provided no evidence that quasar activity is enhanced by major mergers. 
\citet{Mechtley16} examined whether or not quasar hosts exhibit distortions owing to major mergers by comparing 19 quasar host galaxies to 84 inactive galaxies at $z\sim2$. 
These authors found that there was no significant difference in the distortion fractions between the quasar hosts and inactive galaxies, meaning that the triggering mechanism of the quasars was not always a major merger. 
According to \citet{Yang18}, the accretion rates of SMBHs are strongly dependent on the stellar masses of the host galaxies, independently of their sub-Mpc or global $1-10$ Mpc scale environment. 
\citet{Ricarte19} also found that the relation between the black hole accretion rate and star formation rate does not depend on the host galaxy environments, and identified no evidence of a connection between black hole growth and galaxy mergers \citep[but see ][]{Yoon19}.  
\citet{Trakhtenbrot17} determined diverse host galaxy properties of luminous quasars at $z\sim5$ in terms of SMBH fueling mechanisms, suggesting that a galaxy merger may not be a necessary condition for either process, and other mechanisms that are not related to major mergers may drive the rapid growth of the most massive BHs. 
\citet{Ito19} recently found that extreme UV-bright galaxies that are expected to have experienced a starburst phase caused by frequent mergers exist in our  protocluster regions. 

Third, our optical observation might miss some quasars in the rich protoclusters owing to the quasars being in the dust-obscured phase, which appears immediately after experiencing a major merger \citep[][]{Hopkins08}. 
Recently, \citet{Mariko19} stacked the Planck  images of the same HSC protocluster regions as those used in the present study, and found a significant ($>5\sigma$) excess at mid-IR wavelengths, suggesting that  the richest protoclusters at $z\sim4$ may also contain a significant population of obscured AGN. 
Combining these results with our findings appears to be challenging for the merger scenario \citep[e.g.,][]{Hopkins08}, which predicts that quasars are the descendants of obscured AGNs. 
This study combined with that of \citet{Mariko19} simply leads to the inconsistent argument that the progenitors and descendants are not in the same environment, although many uncertainties should be taken into account, such as the quasar duty cycle, and the duration between the major merger and quasar activation. 
Perhaps not all obscured AGNs will evolve into quasars. Moreover, some of them tend to evolve into high-$z$ radio galaxies (HzRGs) in the most overdense regions (the most massive halos) because frequent mergers build up spins of the SMBHs and cause them to launch radio jets \citep[][]{Fanidakis13, Hatch14, Orsi16}. 
The quasar activity, which is maintained by the gas cooling and accretion onto a central SMBH, will be suppressed by radio-mode feedback in a massive halo \citep[e.g.][]{Fanidakis13}. 
As a result, HzRGs are expected to be associated with the richest protoclusters. 
Clustering measurements are absolutely essential for exact interpretations of the major merger scenario. 
We will conduct the clustering analysis of HzRGs and protoclusters based on the HSC-SSP data in forthcoming research.

One should refer to the possible enhancement of quasar activity due to cosmic cold gas. 
Cosmic cold gas inflow in the early universe could also provide a significant amount of gas supply to the SMBHs preferentially in large-scale overdense environments, so that quasar activity could be enhanced in the overdense regions in contrast to our observational results \citep[e.g.,][]{Costa14, Matteo08, Ocvirk08, Sijacki09, Dekel06, Keres05}. 
According to \citet{Dekel06}, cold streams at $z>2$ should still not be fully disrupted in rare massive halo objects with $>10^{12} M_{\odot}$, even if they are diluted by a hot medium heated by quasar feedback. 
However, recent cosmological hydrodynamic simulations \citep[e.g.,][]{Matteo08} demonstrated that strong feedback on massive halos with $>10^{12} M_{\odot}$ at $z\sim6$ stop the cold stream to penetrate into the central BH. 
These authors mentioned that even at $z<6$ in massive halos the feedback energy clearly affects the inflowing cold gas.

\section{CONCLUSION}
We examined the $g$-dropout galaxy overdensity distribution around 570 faint quasar candidates (HSC quasars) at $z\sim4$. 
The possible correlation between the $g$-dropout overdense regions, especially protoclusters and the HSC quasars was examined.   
The protoclusters constructed by \citet{Toshikawa18}  were defined as $g$-dropout galaxy overdense regions with $>4\sigma$ overdensity significance. 
The HSC quasar sample, in which the UV absolute magnitudes were $-26\sim-22$, was constructed by \citet{Akiyama18}. 
We obtained $146$ protoclusters and $570$ quasars in the effective area of $\sim98$ deg$^2$. 

The overdensity distributions in the vicinity of the HSC quasars, 
randomly distributed points over the area, 
and $g$-dropout galaxies were estimated and compared. 
We found that the quasars tend to reside in the same number density as the $g$-dropout galaxies. 
A total of 4 (22) out of the 570, namely $0.7_{-0.4}^{+0.4} (3.9_{-0.8}^{+0.8})$\%, of HSC quasars were found to live in the overdense regions with $>4\sigma$ overdensity significances within 1.8 (3.0) arcmin separation. 
These ratios are almost consistent with the halo mass growth expected from the EPS model. 
Moreover, we found that, among the 570 randomly selected $g$-dropout galaxies, the fraction of those associated with the $>4\sigma$ overdense regions within the $1.8$ (3.0) arcmin separation radius was estimated to be  $0.5_{-0.3}^{+0.3}$\% ($4.4_{-0.9}^{+0.9}$\%). 
These ratios were consistent with the case of the HSC quasars within $1\sigma$ error, suggesting that the HSC quasars do not preferentially appear in the richest protoclusters. 
Combined with the results from the previous study of U18, the number densities of the $g$-dropout galaxies  around the quasars are expected to be independent of their luminosity. 
No preference of quasar activity was found in the $z\sim4$ richest protocluster environment. 
The implications of our results are: 
i) the gas-rich major merger rate is not so high in the protocluster regions, ii) almost all of the high-$z$ quasars may appear through secular processes, or iii) some dust-obscured quasars exist in the protocluster regions. 
The imaging of Hubble Space Telescope combined with spectroscopic observation for the quasar-hosted protoclusters would be key to resolve the possibilities. 


\acknowledgments 

We thank the referee for his/her helpful comments that improved the manuscript.
NK acknowledges support from the JSPS (grant 15H03645). RAO received support from CNPq and the Visiting Scholar Program of the Research Coordination Committee of the National Astronomical Observatory of Japan (NAOJ). 

This work was based on data collected by the Subaru Telescope and retrieved from the HSC data archive system, which is operated by the Subaru Telescope and Astronomy Data Center at the NAOJ. 

The HSC collaboration includes the astronomical communities of Japan and Taiwan, and Princeton University. The HSC instrumentation and software were developed by the NAOJ, the Kavli Institute for the Physics and Mathematics of the Universe (Kavli IPMU), the University of Tokyo, the High Energy Accelerator Research Organization (KEK), the Academia Sinica Institute for Astronomy and Astrophysics in Taiwan (ASIAA), and Princeton University. Funding was provided by the FIRST program from the Japanese Cabinet Office, the Ministry of Education, Culture, Sports, Science and Technology (MEXT), the Japan Society for the Promotion of Science (JSPS), the Japan Science and Technology Agency (JST), the Toray Science Foundation, NAOJ, Kavli IPMU, KEK, ASIAA, and Princeton University. 

This study made use of software developed for the Large Synoptic Survey Telescope (LSST). We thank the LSST Project for making their code available as free software at http://dm.lsst.org.

The Pan-STARRS1 Surveys (PS1) have been made possible through the contributions of the Institute for Astronomy, the University of Hawaii, the Pan-STARRS Project Office, the Max-Planck Society and its participating institutes: the Max Planck Institute for Astronomy, Heidelberg and the Max Planck Institute for Extraterrestrial Physics, Garching, Johns Hopkins University, Durham University, the University of Edinburgh, Queen’s University Belfast, the Harvard-Smithsonian Center for Astrophysics, the Las Cumbres Observatory Global Telescope Network Incorporated, the National Central University of Taiwan, the Space Telescope Science Institute, the National Aeronautics and Space Administration under Grant No. NNX08AR22G, issued through the Planetary Science Division of the NASA Science Mission Directorate, the National Science Foundation under Grant No. AST-1238877, the University of Maryland, and Eotvos Lorand University (ELTE) and the Los Alamos National Laboratory.

\end{document}